\documentclass[12pt]{spieman}  
\usepackage{amsmath,amsfonts,amssymb}
\usepackage{graphicx}
\usepackage{setspace}
\usepackage{tocloft}
\usepackage{lineno}
\usepackage[normalem]{ulem}
\title{False Alarm Rate based Statistical Detection Limit for Astronomical Photon Detectors}

\author[1,3,*]{Albert W.K. Lau}
\author[2,3]{Leo W.H. Fung}
\author[4,5,6,7,8,9]{George F. Smoot}
\affil[1]{Dunlap Institute for Astronomy and Astrophysics, University of Toronto, Toronto, Ontario, M5S 3H4, Canada}
\affil[2]{Institute for Computational Cosmology \& Centre for Extragalactic Astronomy, Durham University, Stockton Rd, Durham DH1 3LE, UK}
\affil[3]{Jockey Club Institute for Advanced Study, Hong Kong University of Science and Technology, Clear Water Bay, Kowloon, Hong Kong}
\affil[4]{Jockey Club Institute for Advanced Study, Hong Kong University of Science and Technology, Clear Water Bay, Kowloon, Hong Kong, \it emeritus}
\affil[5]{Energetic Cosmos Laboratory, Nazarbayev University, Astana, Kazakhstan, \it emeritus}
\affil[6]{Department of Physics, University of California, Berkeley, California, USA, \it emeritus}
\affil[7]{Laboratoire APC-PCCP, Université Sorbonne Paris Cité, Université Paris Diderot, \it emeritus}
\affil[8]{Laboratoire Astroparticule et Cosmologie, Universit{\'e} de Paris, F-75013, Paris, France, \it emeritus}
\affil[9]{Donostia International Physics Center,  University of the Basque  Country  UPV/EHU,  E-48080  San  Sebastian, Spain}

\cftpagenumbersoff{figure}
\cftpagenumbersoff{table} 
\begin{document} 
\maketitle

\begin{abstract}
In ultra-fast astronomical observations featuring fast transients on sub-$\mu$s time scales, the conventional Signal-to-Noise Ratio (SNR) threshold, often fixed at $5\sigma$, becomes inadequate as observational window timescales shorten, leading to unsustainably high False Alarm Rates (FAR).
We provide a basic statistical framework that captures the essential noise generation processes relevant to the analysis of time series data from photon-counting detectors. 
In particular, we establish a protocol of defining detection limits in astronomical photon-counting experiments, such that a FAR-based criterion is preferred over the traditional SNR-based threshold scheme.
We developed statistical models that account for noise sources such as dark counts, sky background, and crosstalk, and establish a probabilistic detection criterion applicable to high-speed detectors. 
The model is testified against the on-site data obtained in the Single-Photon Imager for Nanosecond Astrophysics (SPINA) experiment and consistency is confirmed.
We compare the performance of several detector technologies, including photon-counting CMOS/CCDs, SPADs, SiPMs, and PMTs, in detecting faint astronomical signals. 
These findings offer insights into optimizing detector choice for future ultra-fast astronomical instruments and suggest pathways for improving detection fidelity under rapid observational conditions.
\end{abstract}


{\noindent \footnotesize\textbf{*}Albert Wai Kit Lau,  \linkable{awk.lau@utoronto.ca} }

\begin{spacing}{2}   
\begin{figure}
    \centering
    \includegraphics[width=0.8\linewidth]{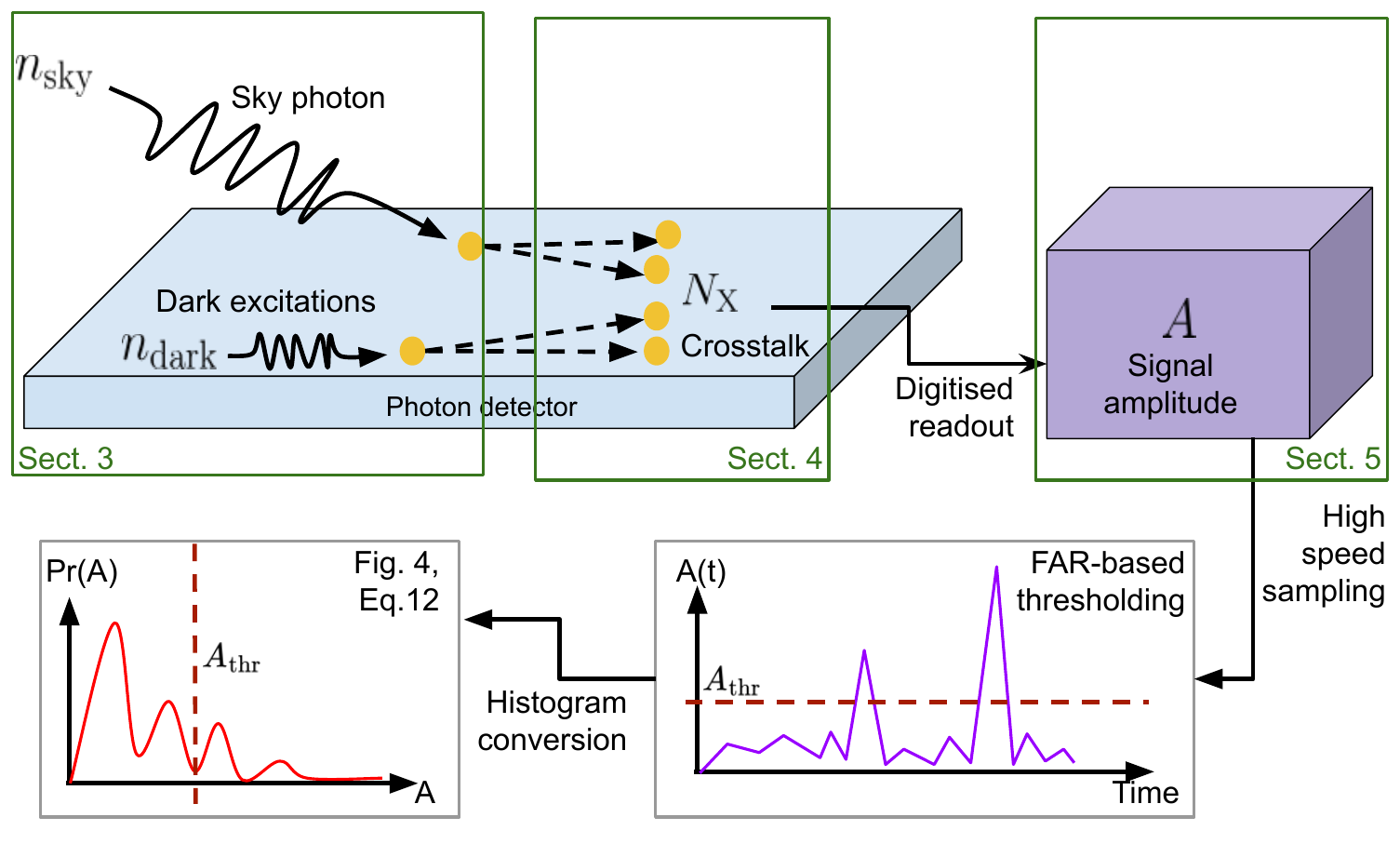}

    \caption{Cartoon illustration of the noise generation process. A more {simplified} and condensed mathematical representation of this process is shown in Fig.~\ref{fig: flowchart}.}
    \label{fig: cartoon}
\end{figure}
\begin{figure}
    \centering
    \includegraphics[width=0.8\linewidth]{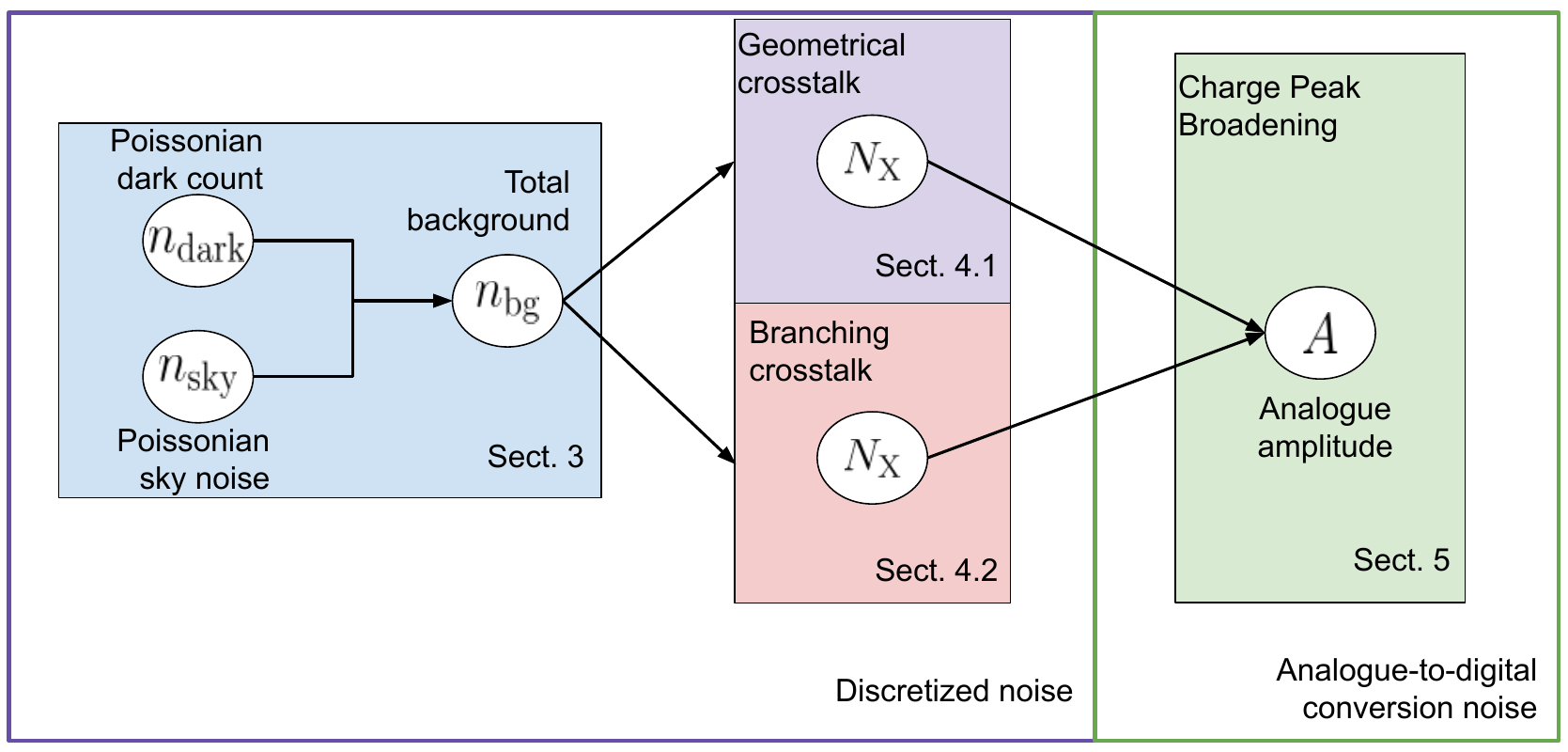}
    \caption{A flowchart of the noise model presented in this work, with an emphasis on how each random variables are probabilistically related. Starting from a model of the total background noise, we gradually introduce 2 models of detector crosstalk. Building on top of this discretized noise-trigger counting model, we finally proceed to provide a more realistic, continuous model that describe the electronic signal received in the detector.}
    \label{fig: flowchart}
\end{figure}

\begin{table}
\centering
\label{tab: dictionary-of-notation}
\begin{tabular}{|| c | c | c ||} 
 \hline
 Notation & Meaning & Units \\
 \hline\hline
 $n_{\rm dark}$ & Dark count noise rate & Counts per second  \\
 \hline
 $n_{\rm sky}$ & Sky background count rate & Counts per second \\
 \hline
 $n_{\rm bg}$ & Total background count rate & Counts per second \\
 \hline
 $\delta t$ & Detector integration time & Seconds \\
 \hline
 $p_{\rm cross}$ & Detector crosstalk probability & \textbackslash \footnote{Unit-less or undefined unit} \\
 \hline
 $\mathrm{FAR}$ & False Alarm Rate & Alarms per second \\
 \hline
 $N_{\rm det}$ & Detector output count & Counts \\
 \hline
 $N_{\rm thr}$ & Alarming threshold count & Counts \\
 \hline
 $N_{\rm X}$ & Crosstalk-affected count & Counts \\
 \hline
 $A_{\rm det}$ & Detector output analogue signal amplitude & ADU \footnote{Arbitrary Detector Output Unit}\\
 \hline
 $A_{\rm thr}$ & Alarming threshold analogue signal amplitude & ADU $^b$ \\
 \hline
 $\gamma_\mathrm{thr, det}$ & Alarming threshold in detected photons & photons \\
 \hline
 $\gamma_\mathrm{thr, in}$ & Alarming threshold in photons input on detector & photons \\
 \hline 
 $R_{\rm n}$ & Read noise of detector system & $e^-$ equivalent\\
 \hline
 $g$ & conversion gain of detector system & ADU $^b$ per photon detected\\
 \hline
 $k$ & Noise correlation factor of detector  & \textbackslash $^a$, [0,1] typical \\
 \hline
 QE & Quantum efficiency of detector  & \% \\
 \hline\hline
 $\langle \cdot \rangle $ & Expectation of random variable $\cdot$ & Same as the random variable $\cdot$ \\
 \hline
 $\mathrm{Pr}(\cdot)$ & Probability mass/density of random variable $\cdot$& \textbackslash $^a$\\
 \hline\hline
\end{tabular}
\caption{A dictionary of all the mathematics notations.}
\end{table}

\section{Introduction}
\subsection{Detection Threshold for Ultra-fast Measurements}
In astronomical observations, the detection criterion is typically defined by a greater than 5 Signal-to-Noise Ratio (SNR), often referred to as the $5\sigma$ criterion. Assuming normally distributed noise, this threshold corresponds to a confidence level of $99.9999713\%$ - equivalently 3 parts in 10 millions, which is generally reliable in most observational contexts. 
However, as we shorten the observational window timescale, especially in the realm of ultra-fast astronomy, this approach becomes less practical due to the increase of false alarms.

Ultra-fast astronomical observations is a emerging technique in astronomy, with focus on timescales ranging from milliseconds to nanoseconds  \cite{li2019program} and targeting transients events \cite{2017NatAs...1..854A, 2019MNRAS.485L.109Z} or transitions\cite{richichi2014final, benbow2019direct}.
While its scientific prospects are promising, it also presents unique challenges that are not relevant in traditional long-exposure optical observations.

For example, continuous observations at a timescale of $1\ ms$ over an entire night (approximately $50,000$ seconds) result in capturing $5 \times 10^7$ frames. 
Using the $5\sigma$ criterion, which corresponds to a false alarm probability of less than $0.0000287\%$, would produce around $10^1$ false alarms per night, identically more than one per hour. 
As the observational timescale decreases to $1\ ns$, the number of captured frames increases to $5 \times 10^{13}$, resulting in overwhelming $10^7$ false alarms per night, or approximately $10^2$ false alarms per second.

This high false alarm rate is clearly unsustainable. The rate of false alarms is directly tied to the observational timescale, making the use of a fixed SNR criterion inadequate for ultra-fast observations. Instead, a more practical approach is to adopt a fixed false alarm rate (FAR). In ultra-fast astronomy, a typical criterion might allow one false alarm per $10^7$ seconds, while other applications may employ thresholds of one false alarm per $10^5$ seconds (approximately equivalent to per night $\approx 5 \times 10^4$ seconds), or even per hundred nights ($\sim 5 \times 10^6$ seconds). In the subsequent sections, we will explore the effectiveness and application of these FAR-based criteria in the context of ultra-fast astronomical detection.
To achieve this, we also developed a noise generation model adequate for describing most generic ultra-fast photon sensors, which can easily be adopted in future experiments of this kind, thus providing useful tools for ultra-fast astronomy.

\subsection{Photon Counting Detectors and Single-Photon Detectors}
Before diving into statistics, we would first review detectors used in astronomical observation.
Conventional optical detectors, such as photographic films and most CCDs, are unable to produce a discernible signal upon detecting a single photon due to high input-referred noise. This noise is typically larger than the signal generated by a single photon (commonly referred to as an $e^-$, as most detectors convert photon absorption into electron excitation). When a sufficient number of photons are detected, the output signal can be approximated by a normal distribution, provided that enough photons are collected.

With advancements in detector technology, modern sensors such as EMCCDs, CCDs, and sCMOS are achieving intrinsic gains or reduced readout noise levels that bring input-referred noise well below $1\ e^-$ \cite{daigle2010darkest, tiffenberg2017single, boukhayma2016noise, khandelwal2024beyond}. At these low noise levels, the output signal deviates from the normal distribution, revealing the quantized nature of photons \cite{teranishi2012required}.

Beyond photon counting, certain detectors are capable of single-photon detection, where individual photons are measured as they arrive. This capability allows for the determination of additional properties, such as photon arrival time and correlations between multiple photons. Single-photon detectors must operate with response and readout times on the order of microseconds or nanoseconds. Technologies such as photomultiplier tubes (PMTs) \cite{iams1935secondary}, silicon photomultipliers (SiPMs) \cite{buzhan2003silicon}, transition-edge sensors (TES) \cite{irwin2005transition}, and microwave kinetic inductance detectors (MKIDs) \cite{mazin2005microwave, shafiee2024characterization} are examples of such systems.
Certain detectors, such as MKIDs, can even discern the energy (or wavelength) of individual photons \cite{mazin2005microwave}. These technologies are evolving into array configurations with imaging capabilities, making them increasingly essential for a wide range of applications in astronomical research. 
These detectors may have different types of noise compared to conventional optical detectors, which we will introduce and formulate in the subsequent sections.

\section{False Alarm Rate based Detection Limit for High Speed Sensors} \label{sect: far-formalism}
\subsection{Defining False Alarms}
We are interested in the astrophysical use case, in which there is a sudden increment in detected photon.
This sudden increment in detected photon can both be driven by stochastic noise fluctuation or physical astronomical phenomena. 
Of scientific interest, we would like to separate between the two scenarios.
In particular, we would like to set an integrated photon count threshold $N_{\rm thr}$ (over some integration time window $\delta t$), so that exceeding the threshold ($> N_{\rm thr}$) would result in a trigger/alarm for follow-up analysis. 
The detection limit in this trigger/alarm based setting can be established by imposing a False Alarm Rate (FAR) limit, which restricts the frequency of noise-triggered alarms.

We first give a concise statistical definition of false alarm, and thus FAR in our experiment setup. 
Suppose at any integration window of size $\delta t$, there is a chance $p_{\rm alarm}$ at which an alarm is triggered. 
Usually, it is by requiring that the integrated signal strength $N_{\rm det}$ at the time window being larger than certain alarm threshold $N_{\rm thr}$.
Across each of the integration windows, we can model the statistical process of checking alarms from all the windows as a binomial experiment.
As such, for a monitoring period of length $T\equiv (1/\mathrm{FAR})$, there are $T/\delta t$ independent binomial trails. 
More explicitly, the probability distribution of the number of alarms $\eta_{\rm alarm}$ is:
\begin{equation} \label{eq: binomial-alarm}
    \mathrm{Pr}(\eta_{\rm alarm}| T, \delta t, p_{\rm alarm}) = \begin{pmatrix}
    T/\delta t \\
    \eta_{\rm alarm}
\end{pmatrix}
p_{\rm alarm}^{\eta_{\rm alarm}}(1-p_{\rm alarm})^{T/\delta t - \eta_{\rm alarm}}
\end{equation}
We wish to adjust the $p_{\rm alarm}$, so that given the false alarm rate FAR, the expected number of alarms during $T$ is exactly unity (or slightly smaller than unity due to quantization error as would be elaborated later):
\begin{equation} \label{eq: far-def}
    \langle \eta_{\rm alarm} \rangle = \frac{1}{\mathrm{FAR}\cdot \delta t} \cdot p_{\rm alarm} \leq 1
\end{equation}

We can now study $p_{\rm alarm}$ as a function of the detection threshold $N_{\rm thr}$ and the noise generation parameters. 
Consider a single-pixel sensor with characteristic parameters (such as dark count rate, sensitivity, and so on, which we will elaborate in the following sections), observing a patch of sky with an averaged background photon rate $\langle n_{\rm bg} \rangle$ over a time bin of $\delta t$, and searching for a photon burst. 
The expected false alarm probability per integration window can be calculated.
This can be translated into an inverse problem of optimizing for $N_{\rm thr}$ given a FAR constraint:

\begin{equation}
\label{eq1}
    p_{\rm alarm} \equiv \mathrm{Pr}(N_{\rm det} \geq N_{\rm thr}|\delta t, \langle n_{sky} \rangle, \langle n_{dark} \rangle, p_{\rm cross}) \leq \mathrm{FAR} \cdot  \delta t
\end{equation}

The above equations are formulated for detecting photon excess; we can similarly derive the statistical equations for detecting photon deficit,
{\ which is simply achieved by modifying the definition of an alarm}
as discussed in Appendix \ref{Appx:detecting photon deficit}.

It is also important to note that the above equations calculate the FAR for a single resolution element of the sky. When the detector has imaging capabilities, the probability must be multiplied by a factor of $\frac{\text{sensor area}}{\text{resolution element size}}$. This implies that the detection threshold increases as the FAR is distributed across the entire sensor area.

{\
\subsection{False Alarm Rate Versus Other Figures-of-Merit}
In this work, we chose to work within the framework of FAR. Indeed, other figures-of-merit (FOM) exist. 
We would argue that FAR is a natural choice in the scientific context of searching for unknown signals and thus deriving the populational characteristics of such signals.

There are indeed various commonly used FOMs in the context of signal detection from noisy data stream.
The most obvious one is to define a notation of power generated by the noise, as known as the noise-equivalent power (NEP) \cite{mackowiak2015nep}.
As such, a notation of the minimal detectable signal power can be defined. 
However, this framework does not naturally come with a way to address the statistical significance of the `signal' as the fluctuations in noise-generated power is not captured in this definition.

Some of the other FOMs are, however, implicitly normalized with respect to some known reference signal.
Following the logic of NEP, if the signal is controllable (as in the laboratory setups, for exampling in medical imaging), the detector quantum efficiency (DQE) can be defined as the ratio of NEP to the minimal power of the injected signal to make it robustly detectable.
Such a notation of minimal detectable signal power is, however, ambiguous when the signal shape is not controllable, as there is no definite way to define the optimal statistical treatment of recovering the signal from the data.
This is in contrast to the case where matched filtering can be defined as the optimal way to recover a known signal from noise, which naturally breaks the ambiguity - as well as offering a statistical assessment of the robustness of the signal detection.

There are no known ultra fast astronomical transients (on the time scale of $\sim \mu s$) at optical wavelengths discovered so far.
As such, the signal is indeed unknown, and we wish to put a (non-)detection constraints on the abundance of these transient events. 
Given this specific scientific use case, we focus exclusively on flagging out the subsets of data which are substantially deviated from noise - without any reference to any signal models.

The use of FAR in this context of population analysis is intuitive. 
Suppose that the alarms trigger much more often than the FAR suggested during a certain observation; it is more likely than not that those alarms are of astrophysical origin.
The number of observed alarms - and its ratio to the (1/FAR) - nails down the statistical significance of the alarms. 
More specifically, one can evaluate the binomial probability of observing the specific number of alarms, given that the expected number of noise-driven alarms is of unity.
The ability to assign probability in such a handy way is one of the strengths of FAR-based metrics.
}

\section{Dark-Count Noises and Sky Backgrounds}
Below, we would introduce a model that captures the noise-generating processes in typical single-photon detectors. 
The physical processes relevant to noise generation are visualized in Fig.~\ref{fig: cartoon}.
With a particular focus in the statistical relation between noise parameters, the hierarchical relations among the random noise variables are shown in Fig.~\ref{fig: flowchart}.

Realistic detectors always exhibit some level of thermal noise. In traditional detectors like CCDs, electrons excited by thermal fluctuations are captured, resulting in dark current, commonly measured in units of $e^-/s$. In single-photon detectors, electrons excited by thermal fluctuations on the detection surface are amplified in the same way as those excited by incident photons, producing output pulses that are indistinguishable from genuine photon events. These false detections are referred to as dark counts and are a primary source of noise in single-photon detectors.

The probability of registering $N_{\rm dark}$ dark counts (or $N_{\rm dark}\ e^-$ dark current) within a time interval $\delta t$ can be modeled by a Poisson distribution \cite{gola2014sipm}:
\begin{equation}
\label{eq2}
\mathrm{Pr}(N_{\rm dark} | \langle n_{\rm dark} \rangle, \delta t) = \frac{(\langle n_{\rm dark} \rangle \delta t)^{N_{\rm dark}}\exp (-\langle n_{\rm dark} \rangle \delta t) }{N_{\rm dark}!} 
\end{equation}

It is important to note that variations in the fabrication process can lead to different regions or pixels of the detector exhibiting different dark count rates, resulting in non-uniform noise characteristics across the detector.

The noise contribution from the sky background, as well as from other background sources, can be modeled similarly to dark counts using the Poisson distribution. However, factors such as the optical system’s efficiency, the collection area, and the detector’s sensitivity must also be considered. For such case we can simply replace $\langle n_{\rm dark} \rangle$ with $\langle n_{\rm bg} \rangle = \langle n_{\rm dark} \rangle + \mathrm{QE}\cdot\langle n_{\rm sky} \rangle$ into the above equation, where QE is the quantum efficiency of the detector.

\section{Crosstalk and Afterpulse}
In some single-photon detectors, a single photon can be mistakenly detected as multiple photons arriving simultaneously, producing an output amplitude several times greater than the expected pulse amplitude from a single photon event. This type of false detection is referred to as crosstalk noise, which originates from residual charge or re-emission during avalanche amplification. Since crosstalk can only occur following the detection of a photon signal or noise event, it is classified as secondary noise \cite{gallego2013modeling}.

Various models exist to describe the probability of crosstalk, though not all provide straightforward analytical solutions. In this context, we introduce two commonly used models: the Geometric crosstalk model and the Branching crosstalk model \cite{gallego2013modeling, vinogradov2012analytical}.

\subsection{Geometric Crosstalk Model}
In the Geometric crosstalk model, a single signal or noise detection event can trigger crosstalk only once. However, the noise generated by crosstalk may in turn trigger additional crosstalk events, making the total number of detected ``photon counts" follow a geometric distribution, When there are $N_{\rm bg} > 1$ photon signal or noise events, the probability of detecting a total of $N_{\rm X}$ counts is given by:

\begin{equation}
    \mathrm{Pr}(N_{\rm X} | N_{\rm bg}) = 
    \begin{pmatrix}
    N_{\rm X}-1 \\
    N_{\rm bg}-1
\end{pmatrix}
(1-p_{\rm cross})^{N_{\rm bg}}(p_{\rm cross})^{N_{\rm X}-N_{\rm bg}}.
\end{equation}

Here, $p_{\rm cross}$ represents the crosstalk probability.

And therefore, given a total background rate $\langle n_{\rm bg} \rangle$ and a fixed integration time $\delta t $, we can derive the distribution of the  marginalized, crosstalk-affected count:
\begin{equation}
\label{eq4}
\begin{split}
    \mathrm{Pr}(N_{\rm X} | \langle n_{\rm bg} \rangle, \delta t) &= \sum_{N_{\rm bg} = 1}^{N_{\rm X}} \mathrm{Pr}(N_{\rm X} | N_{\rm bg}) \mathrm{Pr}( N_{\rm bg} |  \langle n_{\rm bg} \rangle, \delta t ) \\
    &= \sum_{N_{\rm bg} = 1}^{N_{\rm X}} \begin{pmatrix}
    N_{\rm X}-1 \\
    N_{\rm bg}-1
\end{pmatrix}
(1-p_{\rm cross})^{N_{\rm bg}}(p_{\rm cross})^{N_{\rm X}-N_{\rm bg}} \frac{(\langle n_{\rm bg} \rangle \delta t)^{N_{\rm bg}}\exp (-\langle n_{\rm bg} \rangle \delta t) }{N_{\rm bg}!}. 
\end{split}
\end{equation}
Note that the marginalized distribution in (\ref{eq4}) is known as the Polya–Aeppli (geometric Poisson) distribution \cite{nuel2008cumulative}.


\subsection{Branching Crosstalk Model}
In the Branching crosstalk model, a single signal or noise detection event may trigger crosstalk multiple times, and the noise generated from crosstalk can, in turn, trigger further crosstalk events. The crosstalk probability in this model can be described using the Borel–Tanner distribution:


\begin{equation}
    \mathrm{Pr}(N_{\rm X} | N_{\rm bg}) = \frac{N_{\rm bg}}{N_{\rm X}}\frac{(\lambda N_{\rm X})^{N_{\rm X}-N_{\rm bg}}\exp(-\lambda N_{\rm X})}{(N_{\rm X}-N_{\rm bg})!}
\end{equation}

Where $\lambda = -log(1-p_{\rm cross})$. 

Similar to the Geometric model, we can combine the Branching crosstalk model with the event distribution. Assuming the event distribution follows a Poisson process (e.g., primarily driven by dark and background counts), the combined probability forms a generalized Poisson distribution \cite{vinogradov2012analytical}:

\begin{equation}
\begin{split}
    &\mathrm{Pr}(N_{\rm X} | \langle n_{\rm bg} \rangle, \delta t) \\
    &\sum_{N_{\rm bg} = 1}^{N_{\rm X}} \mathrm{Pr}(N_{\rm X} | N_{\rm bg}) \mathrm{Pr}( N_{\rm bg} |  \langle n_{\rm bg} \rangle, \delta t )\\
    &= \frac{\langle n_{\rm bg} \rangle \delta t(\langle n_{\rm bg} \rangle \delta t+\lambda N_{\rm X})^{N_{\rm X}+1} exp(-\langle n_{\rm bg} \rangle \delta t - \lambda N_{\rm X})}{N_{\rm X}!}
\end{split}
\end{equation}

Other types of secondary noise in photon detectors, such as afterpulses and delayed crosstalk, behave similarly to crosstalk. These phenomena are often caused by optical reflections or trapped charge carriers during avalanche detection \cite{nagy2014afterpulse}. Both afterpulses and delayed crosstalk generate a secondary output pulse shortly after the primary one. Since the delay time of the secondary pulse is generally much shorter than the binning time $\delta t$, they can be modeled using the same crosstalk frameworks, where $p_{\rm cross}$ now represents the combined probability of all secondary noise sources, including crosstalk, afterpulses, and delayed crosstalk.

In practice, crosstalk probabilities usually fall between the Geometric and Branching models, with the actual distribution depending on the detector design. Most well-designed single-photon detectors exhibit secondary noise characteristics that closely follow the Geometric crosstalk model \cite{lau2020sky}. Therefore, we will use the Geometric crosstalk model, as described in Eq. \ref{eq4}, for further calculations in this manuscript.

\section{Analogue-to-Digital Conversion and Amplification Noise}
Photon detectors convert incoming photons into an electronic signal, such as charge or voltage, and the readout system converts these into digital numbers or pulses. For realistic detectors' output, we rewrite the FAR equation in terms of signal amplitude $A_{\rm det}$ recorded by the readout system:

\begin{equation}
\label{eq7}
    \mathrm{Pr}(A_{\rm det} \geq A_{\rm thr}|\delta t, n_{sky}, n_{dark}, p_{\rm cross}) \leq \mathrm{FAR} \cdot  \delta t
\end{equation}
The challenge lies in expressing $Pr(A_{\rm det})$ in terms of $Pr(N_{\rm det})$, which we can obtain from the discrete noise model above; using the total probability theorem:

\begin{equation}
\label{eq8}
    \mathrm{Pr}(A_{\rm det}) = \sum_{N_{\rm det}} \mathrm{Pr}(A_{\rm det}|N_{\rm det})\mathrm{Pr}(N_{\rm det})
\end{equation}
Here, $\mathrm{Pr}(A_{\rm det}|N_{\rm det})$ represents the output amplitude distribution given a known photon count detected $N_{\rm det}$, determined by the sensor's characteristics.

In ideal photon-counting or single-photon detectors, the output amplitude is expected to be an integer multiple of the gain, as photon count is quantized and the gain is typically fixed. Therefore, when plotting the output amplitude as a histogram, discrete lines, known as photoelectron (P.E.) peaks, should appear. A 1 P.E. peak corresponds to the output amplitude from a single detected photon, a 2 P.E. peak corresponds to the output from two detected photons, and so forth.

In practice, several factors can prevent the output of a photon detector from being perfectly quantized. These factors include gain variations across different pixels or regions of the detector, photon arrivals during pixel recovery time (particularly in avalanche-based single-photon detectors), and readout noise from the electronics. As a result, the charge spectrum of the output pulse is broadened in realistic photon detectors.

Rather than modeling each factor individually, we can combine these effects into an overall broad statistical distribution. Typically, we use Gaussian distributions for this purpose, as most random factors and electronic noise tend to exhibit Gaussian-like behavior. Thus, the histogram of the output broadens from discrete lines into a mixture of Gaussian distributions, which is commonly referred to as the Gaussian Mixture Model (GMM) \cite{mclachlan2004finite, alvarez2013design}. Using this model, we can express the desired output amplitude distribution given a known photon count detection, $\mathrm{Pr}(A_{\rm det}|N_{\rm det})$.

A normalized Gaussian distribution requires two parameters: the mean ($\mu$) and the standard deviation ($\sigma$). For $N$ different P.E. peaks, $2N$ parameters are required to model the broadening effect. However, we can reduce the number of dependent parameters for GMM fitting by leveraging the inherent properties of the detectors: (1) the linear conversion gain, $g = \mu_1$, and (2) the standard deviation of each Gaussian peak, $\sigma_n$, which grows with the peak number according to an exponential factor $k$ for noise correlation. Typically, $k$ lies between 0 and 1 \cite{collazuol2012sipm}. 
Mathematically, by labelling the n-th peak by $n$ we can express this as $\mu_n = g \cdot n$ and $\sigma_n = \sigma_1 \cdot n^k = R_{\rm n}\cdot g \cdot n^k$ (here we use $R_{\rm n} \cdot g$ to represent $\sigma_1$ for agreement to common convention in CCD/CMOS noise, which we will show). This allows us to reduce the model to just three key parameters: $g$, $R_{\rm n}$, and $k$, which can vary with the detector model and must be determined experimentally. For semiconductor avalanche-based detectors, $k$ is typically around $0.5$  \cite{lau2020sky}, representing the Gaussian randomness inherent in the avalanche process, while vacuum tube detectors like PMT may have $k$ approaching $1$, indicating a worse photon resolving power. 

Thus, the algebric expression for the GMM model is:

\begin{equation}
\label{eq9}
    \mathrm{Pr}(A_{\rm det}|N_{\rm det}) = \frac{1}{\sqrt{2\pi\cdot R_{\rm n} \cdot g \cdot N_{\rm det}^k}}e^{-\frac{(A_{\rm det} - g\cdot N_{\rm det})^2}{2(R_{\rm n}\cdot g \cdot N_{\rm det}^k)^2}}
\end{equation}

In common CCD and CMOS detectors, noise originating from the readout electronics, commonly referred to as read noise ($R_{\rm n}$), arises from the electronics responsible for powering and reading data from the detector. This type of noise is typically independent of integration time. Read noise can also be modeled using the Gaussian Mixture Model (GMM) described above. 

Unlike avalanche-based detectors, where randomness in the avalanche process leads to variable noise, CCD and CMOS detectors exhibit a more consistent behavior since there is no avalanche-related randomness. As a result, the read noise remains constant, regardless of the number of photons collected (i.e., the peak number). Therefore, we can simplify the model by setting the exponential growth factor $k=0$, which implies that the standard deviation remains constant across all peaks. Traditionally, this read noise $R_{\rm n}$ of CCD/CMOS is reported in the unit of $e^-$ equivalent, i.e. electrons excited by photon detected. Following this convention, we express the GMM variance as $\sigma_1 = \sigma_2 = \dots = \sigma_N = R_{\rm n} \cdot g$, together with the conversion gain $g$ of the detector.

\section{Detection Limit Simulation}
To quantify the detection limits of high-speed astronomical detectors, we integrate the statistical models developed in the previous sections into a numerical simulation. The goal is to evaluate the performance of various detectors under different observational conditions, particularly focusing on their ability to detect faint astronomical signals over ultra-fast timescales.

Here, we wish to calculate the distributional properties of the detector signal amplitude $A_{\rm det}$ given a pre-determined background noise rate $\langle n_{\rm bg}\rangle$ and a window size $\delta t$. 
This distribution would provide insight in quantifying the amount of false event triggers driven solely by various noise.
Using the equations derived earlier, we compute the detection threshold for each detector by propagating the signal amplitude distribution, incorporating both intrinsic detector noise and external noise sources. To capture the combined effects of dark counts, background noise, and crosstalk across different detector types, we employ the formulations for $\mathrm{Pr}(N_{\rm dark} | \langle n_{\rm dark} \rangle, \delta t)$, $\mathrm{Pr}(N_{\rm X} | N_{\rm bg})$ and $\mathrm{Pr}(A_{\rm det}|N_{\rm det})$ from Eq. \ref{eq2}, \ref{eq4}, and \ref{eq9}, respectively. These terms are then propagated to obtain the overall probability density of the detector's output amplitude by combining Eq. \ref{eq4} and \ref{eq8}:
\begin{equation}
\label{eq: marginalise-adu-prob}
\begin{split}
    &\Pr(A_{\rm det}|\langle n_{\rm bg} \rangle, \delta t) =\\ 
    &\sum^{\infty}_{N_{\rm det}} \sum^{N_{\rm det}}_{N_{\rm bg}=1} \mathrm{Pr}(A_{\rm det}|N_{\rm det}) \mathrm{Pr}(N_{\rm det} | N_{\rm bg}) \mathrm{Pr}(N_{\rm bg} | \langle n_{\rm bg} \rangle, \delta t)
\end{split}
\end{equation}
Finally, we compare this probability density of the detector's output amplitude with the desired false alarm rate, as defined by Eq. \ref{eq7}, to determine the detection threshold $A_{\rm thr}$.

Before evaluating the distribution of $A_{\rm det}$, we would like to introduce a few quantities that are re-scaled from $A_{\rm det}$, with the purpose of comparing the performance between different detector technologies on equal footing.
The detection threshold $A_{\rm thr}$ can be expressed in terms of the detector’s conversion gain $g$, which allows us to compute the corresponding photon count threshold for detected signals:

\begin{equation}
\gamma_\mathrm{thr, det} = \left\lceil\frac{A_{\rm thr}}{g}\right\rceil
\end{equation}

Here, $\gamma_{\text{thr, det}}$ represents the minimum number of photons that must be detected to surpass the noise floor. To facilitate a comparison between different detector technologies, we further normalize the detection threshold by the detector's quantum efficiency (QE), yielding:

\begin{equation}
\gamma_\mathrm{thr, in} = \left\lceil\frac{A_{\rm thr}}{g }\right\rceil \frac{1}{\mathrm{QE}}
\end{equation}

This normalization accounts for variations in the quantum efficiencies and conversion gains across detectors, enabling a fair comparison of their intrinsic performance.

{\ We implemented a simplified demo simulation using a Python-based framework to model and compare several commonly used detectors in astronomical observations, with the purpose of highlighting the how various properties on different detector affect their prospect in detecting ultra fast transients.} 

The key detector types evaluated in the simulation are:

\begin{itemize}
    \item Photon-counting CMOS/CCD detectors \cite{khandelwal2024beyond}
    \item Single-Photon Avalanche Diodes (SPAD) \cite{naletto2009iqueye}
    \item Silicon Photomultipliers (SiPM) \cite{lau2022development, lau2023initial}
    \item Photomultiplier Tubes (PMT) \cite{lehmann2008performance}
\end{itemize}

{\ For fair comparsion, we consider parameters of commercially available single-pixel models of SPAD, SiPM and PMT, where ideal optics match a star's point spread function onto the detector; while for the CMOS/CCD, we assume the same point spread function is projected onto a single pixel. For detailed comparison, there are factors beyond the sensor (e.g. coupling to front end optics) and should be analyzed on system basis.}
For the simulation, we assume a typical sky background rate of $\langle n_{\rm sky} \rangle = 100$ photons per second onto the detector. The typical detector parameters are shown in Table \ref{fig:detector parameters}. 

\begin{table}
\centering
\caption{Detector Parameters for Simulation}
\label{fig:detector parameters}
\begin{tabular}{||c | c c c c||} 
 \hline
 Detector parameters & photon counting CMOS/CCD & SPAD & SiPM & PMT \\ [0.5ex] 
 \hline\hline
 $n_{\rm dark}$ (Counts per second) & $0.01$ & $100$ & $1000$ & $100$ \\ 
 \hline
 $p_{\rm cross}$ & $0$ & $0.01$ & $0.1$ & $0.03$ \\
 \hline
 $g$ ($ADU$\footnote[3]{Here the ADU chosen is in fact $e^-$ after amplification} per photon detected) & $10$ & $10^4$ & $10^6$ & $10^5$ \\
 \hline
 $R_{\rm n}$ ($e^-$ Eqv.)& $0.3$ & $0$ & $0.1$ &$0.5$\\
 \hline
 k & $0$ & $0$ & $0.5$ & $0.5$\\
 \hline
 QE (\%) & $85$ & $50$ & $35$ & $20$ \\
 \hline
\end{tabular}
\end{table}

We also assume that there is no lower limit on the frame time for the detectors, though this is not realistic in practical applications. Photon detectors like SiPM or PMT can operate on nanosecond timescales, while CMOS/CCD detectors typically have a frame rate limit of around 1,000 frames per second (fps).

\begin{figure}[!ht]
    \centering    
    \includegraphics[width=0.9\columnwidth]{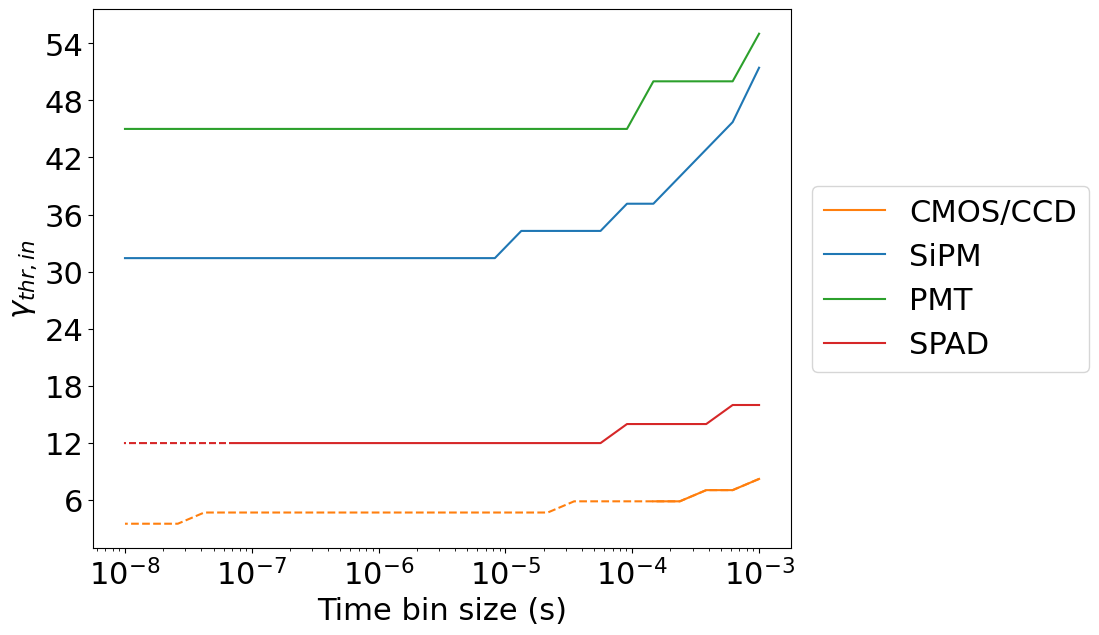}
    \caption{Detection limit of detectors in photon counts arriving at the detector surface $\gamma_{\rm thr,in}$, as a function of the time bin/window size $\delta t$. Note however that the performance of photon counting CMOS/CCD should be downgraded to account for the slow frame rate and inter-frame dead time, effectively making the continuous ultra-fast monitoring unsustainable on those devices {\ (indicated with broken line). Similarly, dead time of SPAD devices lead to reduced performance below hundred nanoseconds (indicated with broken line). Practically the broken lines shall be understood as unrealistic extrapolations.}}
    \label{fig:limit_count}
\end{figure}

Using the detector specifications in Table.~\ref{fig:detector parameters}, we determine the detection threshold by considering a fixed FAR budget of $\mathrm{FAR} =  10^{-7}s^{-1}$, equivalently one per $10^7$ seconds, and express the threshold for each detectors in terms of $\gamma_{\rm thr,in}$. 
Fig. \ref{fig:limit_count} illustrates the detection limits of these detectors in terms of photon counts arriving at the detector surface (equivalently $\gamma_{\rm thr,in}$). {\ As shown in this figure, the CMOS/CCD detector serves as a reference. Despite its superior sensitivity attributed to higher quantum efficiency, lower dark noise, and absence of crosstalk, its limited frame rate and significant inter-frame dead time render it unsuitable for continuous ultra-fast observations.}

{\ SPAD detectors also seems to demonstrate good performance in the ultra-fast regime, but they suffer from dead times on the hundreds of nanosecond to microsecond scale, limiting their ability to detect multiple photons arriving simultaneously; this limitation can be mitigated by deploying multiple SPAD detectors in parallel \cite{capraro2009first}. Meanwhile, SiPM and PMT detectors exhibit higher detection limits but can resolve multiple photons within a few nanoseconds \cite{Seitz2017SiPM}. Among these, SiPM detectors perform slightly better because their photon-resolving capability does not degrade as rapidly with concurrent photon arrivals compared to PMTs.}

\section{Application on Single-Photon Imager for Nanosecond Astrophysics (SPINA) Data}
\begin{figure}[!ht]
    \centering
    \includegraphics[width=0.9\columnwidth]{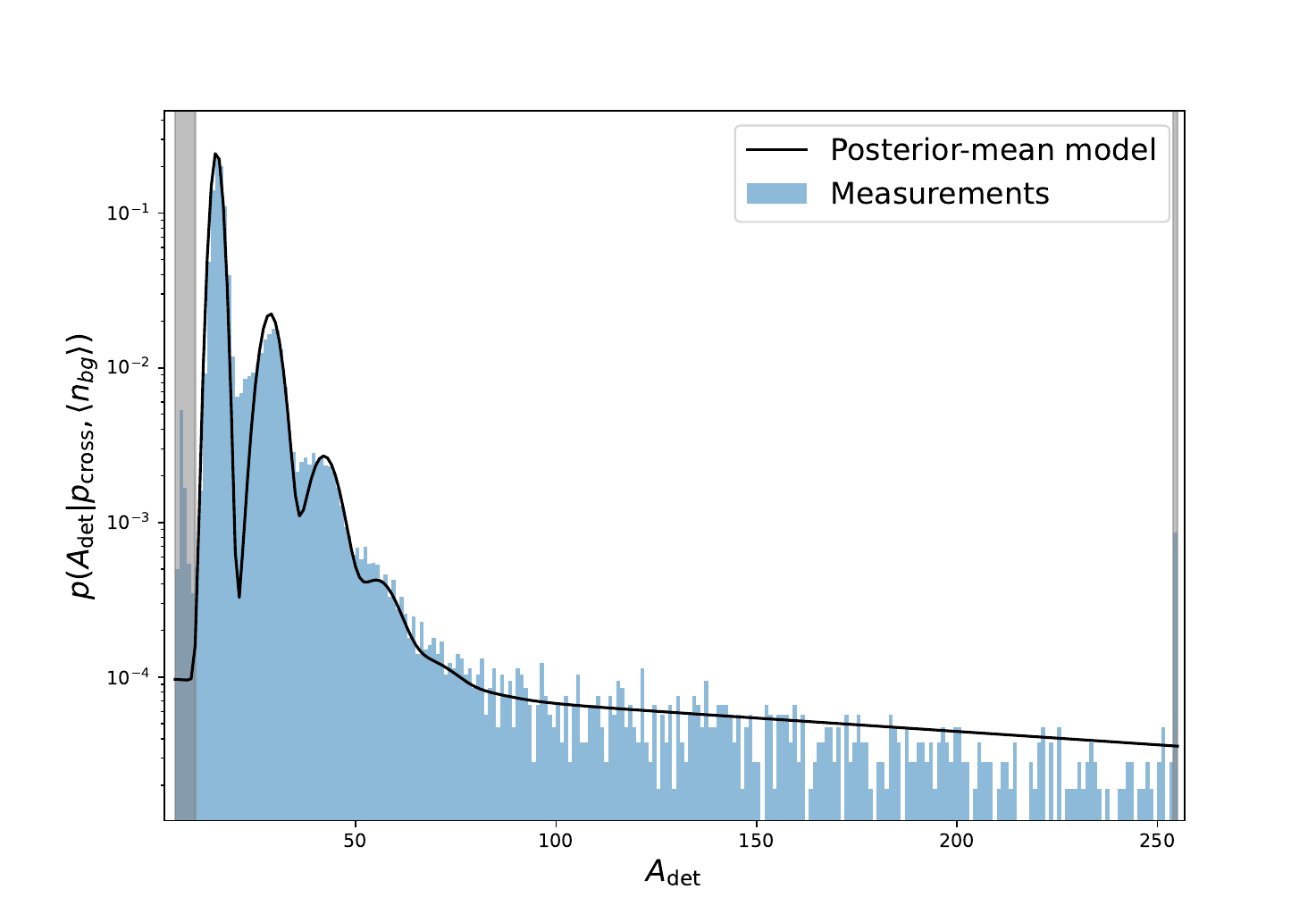}
    \caption{Comparison of the measured SPINA dark frames with the noise model predictions at posterior mean. The grey bands at both ends of the ADU spectrum indicates the masked data during the fitting process, as those features are not captured by our noise generation model. }
    \label{fig: posterior-mean-vs-data}
\end{figure}
We would like to testify if the above noise generation model can indeed explain realistic astronomical single photon detectors.
For such purpose, we applied our model to the data collected in the Single-Photon Imager for Nanosecond Astrophysics (SPINA) experiment.
The SPINA system is an experimental setup developed in \cite{lau2022development}, thereby a SiPM based detector is developed for astronomical purpose.
The SiPM is temperature controlled with cryostats, and a FPGA pipeline has been developed for achieving ultra-fast readout at GHz rate, thus enabling continuous photon arrival monitoring at the time resolution of $\delta t = 8 $ns.
The system had been sent to the NUTTellA-TAO observatory in Kazakhstan for on-sky testing in 2023 \cite{lau2023initial}, providing valuable data for the verification of technologies and feedback for the scheduled upgrade.
In particular, we would use the dark frames obtained on-site to verify if our noise generation model described in this paper is realistic. 

We took the dark frame of SPINA spanning a total duration of $81.8$ s, equivalently consisting of $1.02 \times 10^{10}$ samples. 
Due to such ultra-fast sampling rate and consequently the giantic amount of data produced, we deployed a FPGA based data selection/compression algorithm.
A data selection threshold was chosen, so that only samples with $A_{\rm det} > 5$ are recorded, where $8$-bits are assigned to define the analog-to-digital units (ADU).
This aggressive, yet effective data cut shrinks the data size by a factor of $10^5$ specifically for this dark frame.
The samples are further packed together with time stamps, and the estimated x-y positions of the signal across the detector plane.
Due to this data selection process and other mild systematics, additional treatments are applied to fairly compare the data with our noise generation model.
The details of the data fitting process can be found in Appendix.~\ref{appx: spina-likelihood}.

In essence, we demonstrated satisfactory agreement between our noise generation model and the actual observed patterns in the SPINA dark frame.
The fitting results are visualized in Fig.~\ref{fig: posterior-mean-vs-data}.
System-specific corrections are applied on top of our noise generation model presented throughout the preceding sections, although the amplitude of such correction is as small as one part in $10^{6}$ compared to the main noise generation model.
This demonstrates that our noise generation model are indeed applicable to real detectors.

\section{Combining Multiple Detectors for Coincidence Detection}
From the preceding analysis, we can see that single-photon detectors suffer from excess false photons due to secondary noises such as crosstalk. 
These secondary noises are correlated with the primary noise and with each other, introducing some degree of temporal ``bunching"; that is, they tend to produce short but intense noise signals that mimic the astronomical transients we aim to detect.

To achieve a lower detection limit, we propose using multiple detectors and employing coincidence detection. Since the noise from different detectors is uncorrelated, we can distinguish real transients from secondary noises. This can be realized through electronically isolated detectors with beam splitters or through independent detector and telescope systems located far apart. Completely separate and distant detectors can further enhance sensitivity, reduce false alarm rates (FAR), and mitigate errors from local noise sources such as lightning, man-made light flashes, or even satellite flares from low Earth orbit, provided the observation stations are sufficiently distant.

{\
\subsection{Defining Multiple Detectors Alarms}
There are a lot of existing algorithms for combining data from multiple detectors to mitigate noise \cite{countung-stat,counting-stat2}.
Those algorithms mostly focus on applications in particle physics experiments, which addresses a slightly different experiment setup from ours presented in this article. 
For such a reason, we would elucidate the technical details of defining multi-detectors alarm in the following subsection.

We have defined the FAR formalism of a single detector in Section.~\ref{sect: far-formalism}.
However, there can be ambiguities in the definition of alarming in the context of combining the data stream from multiple detectors.
Suppose there are $M$ detectors, each of them is governed by independent noise generation process with their own noise parameters inherently determined by the fabrication error in each of the detectors.
At a given integration window $\delta t$, each of the $M$ detectors would therefore produces alarms at the probability of $p_{\rm alarm,(i)}$ where $i$ runs over the detectors.
For the ease of discussion, we would assume that the alarming probabilities of different detectors $p_{\rm alarm}$ are identical, thus dropping the $(i)$ subscript.
A multi-detectors alarm is defined as \textbf{the simultaneous alarm in all detectors at exactly the same integration window.}
As the noise generation process on each of the detectors are independent, the multi-detectors alarming probability is simply $p_M = (p_{\rm alarm})^{M}$. 

This definition is generic. In the case where there are optical path difference (which is deterministic upon calibration) across the detectors, it is possible to apply time-shifting to the data streams so that the time delay is cancelled.
As such, the definition of `exactly the same integration window' still applies in the generic case.

\subsection{Enhancing Sensitivity with Multiple Detectors}
The multi-detectors FAR can now be defined after defining a multi-detectors alarm.
The procedure of flagging out alarm with the new alarm probability $p_M$ is the same as before: it is a binomial experiment as argued in the discussions around Eq.~\ref{eq: binomial-alarm}.
The multi-detectors FAR is defined analogously as Eq.~\ref{eq: far-def} with replacement of $p_{\rm alarm}$ by $p_M$.
}

Following Eq. \ref{eq7}, we can formulate the new detection threshold for $M$ detectors in coincidence observation as 

\begin{equation}
    p_M \equiv \left[\mathrm{Pr}(A_{\rm det} \geq A_{\rm thr}|\delta t, n_{sky}, n_{dark}, p_{\rm cross})\right]^M \leq \mathrm{FAR} \cdot  \delta t
\end{equation}

To demonstrate the effect of coincidence detection with multiple detectors, we reran the above simulation with 1, 2, and 3 SiPM, PMT or SPAD detectors with signal splitted (CMOS/CCD detectors excluded as they are not capable for fast coincidence detection).

\begin{figure}[!ht]
    \centering
    \includegraphics[width=0.9\columnwidth]{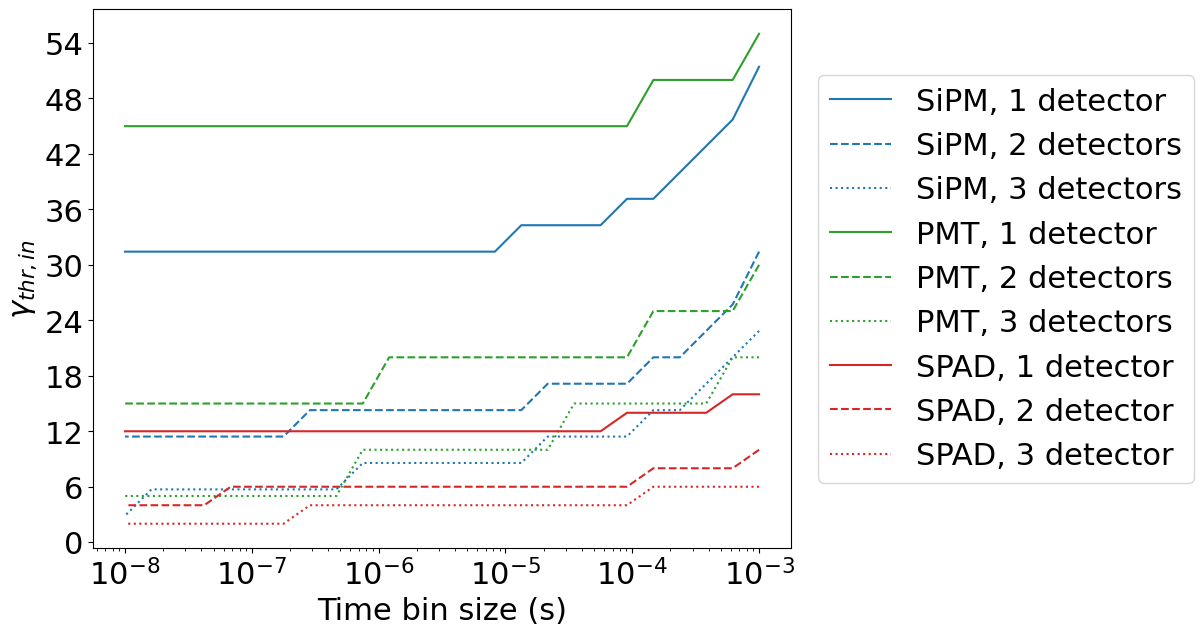}
    \caption{Detection limit of detectors in photon counts arriving at each of the detector surface $\gamma_{\rm thr,in}$, as a function of the time bin/window size $\delta t$, with multiple detector counts in coincidence mode.}
    \label{fig:photon_limit_3_detectors}
\end{figure}

From Fig. \ref{fig:photon_limit_3_detectors}, we can see that the detection limit of coincident single photon detectors roughly improves by $M$-fold ($M$: number of detectors) in longer timescales of $\sim 1$ ms, as expected for a sky-background or dark noise dominated setup. Since the setup is based on a beam splitter, the effective collection area will also decrease by $M$-fold, indicating that there will be no gain in sensitivity regardless of how many detectors are used.

In contrast, we observe a substantial improvement when the integration time shortens to the nanosecond scale. The higher the time resolution of the coincidence detector, the easier it is to discriminate between ``chance" coincidences and true signals. For example on timescales of $10ns$ on SiPM detectors, the detection limit improves from 31 photons (1 detector) to 12 photons (2 detectors), or even 3 photons (3 detectors) arriving at the detector surface. This demonstrates the power of coincidence detection in eliminating the secondary noise that dominates single-photon detectors. Note that we set the quantum efficiency of the SiPM to be 35\%, essentially meaning that only one photon detected by each of the three SiPMs is sufficient to suppress the FAR to below $1\times10^{-7}$.

\section{Conclusion}

This work introduces a statistically driven methodology for determining detection limits in photon detectors used in ultra-fast astronomical observations. By shifting from a traditional Signal-to-Noise Ratio (SNR) threshold to a False Alarm Rate (FAR)-based detection criterion, we address the challenge of managing false positives that arise from reducing observational timescales. The FAR-based model, grounded in Poisson statistics and generalized to account for dark counts, sky background, and crosstalk noise, provides a more reliable approach for high-speed astronomical detection. Our statistical framework was applied to different detector technologies, and simulations revealed significant differences in performance. This approach can be generalized to evaluate future detector technologies and observational strategies, ensuring that false alarm rates are controlled while maintaining the sensitivity required for detecting transient astronomical events.

\appendix

\section{Remarks on the Detection Limit for detecting photon deficit}
\label{Appx:detecting photon deficit}
In the main text of this work, we exclusively focus on the use case, in which there is a ultra-fast astronomical transient that suddenly brightens up, thus delivering an unusual increment of photons counts over a very short time window. 

As a complement, there is also another type of ultra-fast astronomical transients, that instead of a sudden brightening, a sudden dimming is observed.
A more familiar astrophysical phenomenon of this type (on a longer time scale) is the observation of exoplanets transit.
Hereafter, we would refer to this type of events as photon deficit.

The detection limit formulation for detecting photon deficit is simply:
\begin{equation}
    \Pr(N_{\rm det} \leq N_{\rm thr, -}|\delta t, n_{sky}, n_{dark}, p_{\rm cross}) \leq \mathrm{FAR} \cdot  \delta t, 
\end{equation}
and
\begin{equation}
    \Pr(A_{\rm det} \leq A_{\rm thr, -}|\delta t, n_{sky}, n_{dark}, p_{\rm cross}) \leq \mathrm{FAR} \cdot  \delta t.
\end{equation}
However, these criteria are rarely used. As the bin time shortens, the expected number of photons per bin decreases significantly, and the detection threshold may fall below zero, leading to a non-physical detection limit. For example, with a background flux of $1 \times 10^6$ counts per second and a bin size of $1\ ns$, it is likely that most bins will capture no photons at all, causing the detection threshold to drop to non-physical value.

\section{Statistical Analysis of SPINA data}
\label{appx: spina-likelihood}
In the main text, we focused on a generic model applicable to various types of sensors, thus omitting device-specific systematics.
In particular, there are some known artifacts in SiPM that are not captured in the generic model. 
As these artifacts only slightly change the prediction from the generic model, we would leave them detached from our generic model.
We would like to compare our noise generation model with the real data, more specifically with the dark frames taken by the \textbf{Single Photon Imager for Nanosecond Astrophysics} (SPINA) experiment.
For such a purpose, we would include these additional detector-specific behaviors in the modeling.
Here we outline how the noise parameters can be estimated from the real data.

The inter-samples time of our SPINA system is $\delta t = 8$ ns (including the sensor dead time), and we logged the output of the sensor in terms of ADU. 
We implemented a data compression scheme; as such, the ADU measurements with ADU $<5$ were discarded.
The recorded data stream is thus a sequence of ADU values in which each ADU value being attached with a time stamp of $\delta t = 8$ ns resolution.
Because of the data filtering scheme, there is substantial selection bias involved in the recorded data stream, thus preventing direct comparison with our noise model.

To mitigate the aforementioned selection bias, we consider the fraction of data points that were discarded during the observation. 
Suppose we measured the dark for an extended period of time $T$, and we collected $N_{\rm A}$ ADU values, which implies the duty cycle $f_{\rm duty}$ of the system is:
\begin{equation}
    f_{\rm duty} = \frac{N_{\rm A}\delta t}{T},
\end{equation}
so that the duty cycle occupied by the dark should ideally be as small as possible. 
For SPINA system, we reported $f_{\rm duty} \approx 10^{-5}$. 
Obviously, the duty cycle depends on the noise generation parameters.
Given the noise generation rate $\langle n_{\rm bg}\rangle$, the non-detection probability is:
\begin{equation} \label{eq: non-detection-prob}
    p_{\rm non} = \mathrm{Pr}(N_{\rm bg}=0 | \langle n_{\rm bg}\rangle, \delta t )
\end{equation}
If we assume the lack of correlation in the dark count, the probability of logging $N_{\rm A}$ ADU values out of the $T/\delta t$ frames is a binomial distribution:
\begin{equation}
\begin{split}
    \mathrm{Pr}(N_{\rm A} | \langle n_{\rm bg}\rangle, \delta t ) &= \begin{pmatrix}
    T/\delta t \\
    N_{\rm A}
    \end{pmatrix} p_{\rm non}^{(T/\delta t - N_{\rm A})}(1-p_{\rm non})^{N_{\rm A}} \\
    &\rightarrow \mathrm{Gaussian}\left(\frac{T(1-p_{\rm non})}{\delta t},  \frac{T(1-p_{\rm non})p_{\rm non}}{\delta t}\right) ,
\end{split}
\end{equation}
where on the second line we use the well known convergence properties of Binomial distribution to speed up numerical evaluation.
Denote the measured data stream as $D \equiv \lbrace A_{\rm det, i} | i=1,...,N_{\rm A}\rbrace$, and more compactly the model parameters as $\Theta \equiv \lbrace \langle n_{\rm bg}\rangle , p_{\rm cross}, A_1, \sigma, k \rbrace$ the likelihood in which the data is generated by some specific noise parameter is thus:
\begin{equation}
    \mathrm{Pr}(D|\Theta ; \delta t) = \mathrm{Pr}(N_{\rm A} | \langle n_{\rm bg}\rangle, \delta t )\prod_i^{N_{\rm A}} \mathrm{Pr}(A_{\rm det, i} | \Theta),
\end{equation}
The terms in the series of products are modeled as identical and independent realisation of noise, following from Eq.~\ref{eq: marginalise-adu-prob}.
There is however a slight modification from the model shown in the main text. 
We found in the data that instead of a Gaussian-Mixture model with linearly spaced Gaussian peak (which scales as $g \cdot N_{\rm det}$), we observe affinity in peak spacing. 
More specifically, the peak spacing scales as $(A_{1} \cdot N_{\rm det} + 2)$, where $A_1$ is a model parameter to be determined.
Such an additional constant offset among the peak separations can be attributed to the definition of the ADU zero point.
We have incorporated this finding when fitting to the data.
Apart from the quantized dark count noise modeled in Eq.~\ref{eq: marginalise-adu-prob}, we also find a smooth component of noise that spans the entire ADU spectrum. 
We hypothesise that this is due to residual radio-frequency interference (RFI) in our signaling system. 
We model this RFI component by a simple exponential model parametrized by the logarithmic e-folding scale $\Delta_\texttt{a}$: 
\begin{equation}
    \mathrm{Pr}(A_{\rm det, i}|\mathrm{RFI}, \Delta_a) \propto \exp( -A_{\rm det, i}/10^{\Delta_\texttt{a}} ),
\end{equation}
the proportionality constant is uniquely determined by the normalisation condition.
Suppose a fraction of $p_{\rm RFI}$ of the measured ADU values are contaminated by this smooth spectrum instead.
As such, the observed ADUs are explained in this composite model:
\begin{equation}
    \mathrm{Pr}(A_{\rm det, i}|\Theta, p_{\rm RFI}, \Delta_\texttt{a}) = \mathrm{Pr}(A_{\rm det, i}|\Theta, \mathrm{Dark})(1-p_{\rm RFI}) + \mathrm{Pr}(A_{\rm det, i}|\mathrm{RFI}, \Delta_\texttt{a})p_{\rm RFI}.
\end{equation}
Note however that the non-detection probability $p_{\rm non}$ as modeled in Eq.~\ref{eq: non-detection-prob} also received an extra correction factor from the RFI component. 
So the resulting total non-detection probability is $p_{\rm non}' = p_{\rm non}(1-p_{\rm RFI})$. 

Using this likelihood, we can infer the posterior of the noise generation model given the data $D$. 
By invoking the Bayes' theorem, we obtain:
\begin{equation}
    \mathrm{Pr}(\Theta, p_{\rm RFI}, \Delta_\texttt{a} | D; \delta t) \propto \mathrm{Pr}(D|\Theta, p_{\rm RFI}, \Delta_\texttt{a} ; \delta t) \mathrm{Pr}(\Theta, p_{\rm RFI}, \Delta_\texttt{a}) ,
\end{equation}
and we adopt an uniform prior for all our model parameters.
The posterior is evaluated using nested MCMC sampling algorithm implemented in the package \textsc{dynesty} \cite{dynesty}. 

\begin{figure}[!ht]
    \centering
    \includegraphics[width=0.9\columnwidth]{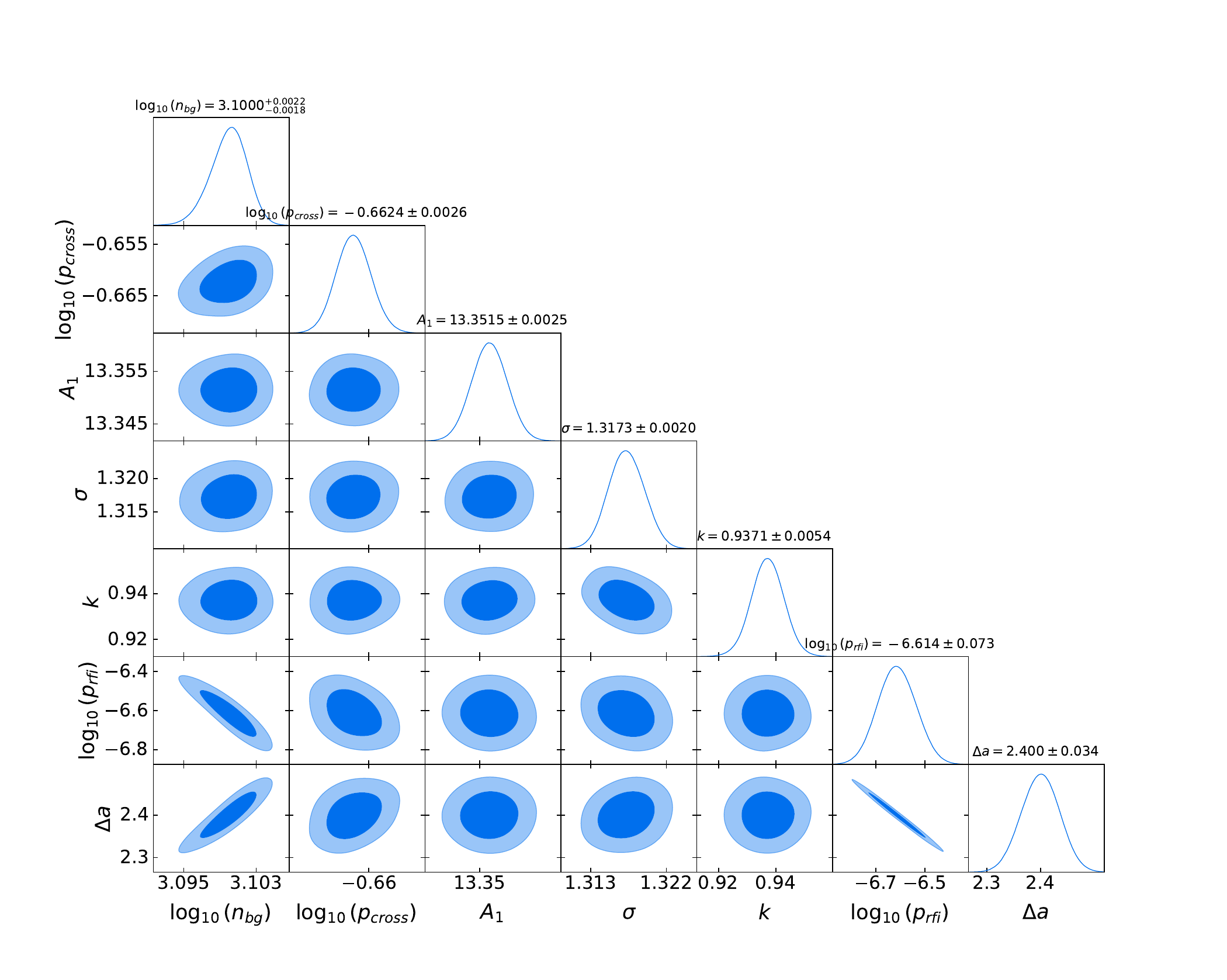}
    \caption{Estimated noise model parameters using the SPINA dark frame measured on-site. }
    \label{fig:spina-posterior-contours}
\end{figure}

We found that we did indeed shield the signaling system from RFI relatively well, constraining $p_{\rm RFI} \approx 10^{-6}$, suggesting that such a component of noise is subdominant in our noise modeling.
We show the posterior of the noise parameters in Fig.\ref{fig:spina-posterior-contours}, and contrast the prediction of the posterior mean with the data in Fig.\ref{fig: posterior-mean-vs-data}.

\subsection*{Disclosures}
The authors declare that there are no financial interests, commercial affiliations, or other potential conflicts of interest that could have influenced the objectivity of this research or the writing of this paper.

\subsection* {Code, Data, and Materials Availability} 

The experimental data will be shared on reasonable request to the corresponding author.

\subsection* {Acknowledgments}
Albert Wai Kit Lau is a Dunlap postdoctoral fellow; The Dunlap Institute is funded through an endowment established by the David Dunlap family and the University of Toronto.

Leo Fung acknowledges funding from the UK Science and Technology Facilities Council via grant ST/X001075/1.

We acknowledge support from the HKUST Jockey Club Institute for Advanced Study (IAS), Hong Kong University of Science and Technology, the Energetic Cosmos Laboratory, Nazarbayev University, the Assy-Turgen Observatory and the Fesenkov Astrophysical Institute for the SPINA data.


\bibliography{reference,ref-stat, additional_ref}   
\bibliographystyle{spiejour}   


\vspace{2ex}\noindent\textbf{Albert W.K. Lau} is a Dunlap Postdoctoral Fellow at the University of Toronto. He received his PhD degree in physics from Hong Kong University of Science and Technology in 2023, under the supervision of George Smoot. He has research interests in astronomical instrumentation and time-domain observation. His research focuses on developing cutting-edge detectors and electronic systems for transient astronomy, spanning optical and radio wavelengths.

\vspace{2ex}\noindent\textbf{Leo W.H. Fung} is a postdoctoral research associate at Durham University. 
He received his BEng degree in Computer Science from The University of Hong Kong in 2019 and PhD degree in physics from Hong Kong University of Science and Technology in 2024, under the supervision of George Smoot.
He has broad research interests in astronomy, including the development of new data pipeline and statistical technique for exploring the nature of dark matter from different types of observations. 

\vspace{2ex}\noindent\textbf{George F. Smoot} is a Professor Emeritus at University of California, Berkeley, and holds various positions in a few institutions. 
He was awarded with the Nobel Prize in Physics in 2006 for his contributions in leading the Cosmic Background Explorer (COBE) experiment, which measured for the first time, the inhomogeneities of the primordial Universe.  
He received his PhD degree in Physics from Massachusetts Institute of Technology in 1970.

\listoffigures
\listoftables

\end{spacing}
\end{document}